\documentstyle[aps,eqsecnum]{revtex}
\draft

\newcommand{\e}{{\rm e}}
\def\q{\quad}
\def\d{\dot}
\def\dd{\ddot}

\begin{document}

\title{              Scaling Solutions in
                    Robertson-Walker Spacetimes}

\author{                R. J.  van den Hoogen}

\address{       Department of Mathematics, Statistics
                                and Computer Science,\\
                Saint Francis Xavier University,
                Antigonish, Nova Scotia, B2G 2W5}

\author{                A. A. Coley}

\address{       Department of Mathematics, Statistics
                                and Computing Science\\
                Dalhousie University,
                Halifax, Nova Scotia B3H 3J5}

\author{                D. Wands}

\address{       School of Computer Science and Mathematics\\
                University of Portsmouth,
                Portsmouth, PO1 2EG, U.K.}

\maketitle

\begin{abstract}
We investigate the stability of cosmological scaling solutions
describing a barotropic fluid with $p=(\gamma-1)\rho$ and a
non-interacting scalar field $\phi$ with an exponential potential
$V(\phi)=V_0\e^{-\kappa\phi}$. We study homogeneous and isotropic
spacetimes with non-zero spatial curvature and find three possible
asymptotic future attractors in an ever-expanding universe. One is the
zero-curvature power-law inflation solution where $\Omega_\phi=1$
($\gamma<2/3,\kappa^2<3\gamma$ and $\gamma>2/3,\kappa^2<2$). Another
is the zero-curvature scaling solution, first identified by Wetterich,
where the energy density of the scalar field is proportional to that
of matter with $\Omega_\phi=3\gamma/\kappa^2$
($\gamma<2/3,\kappa^2>3\gamma$).  We find that this matter scaling
solution is unstable to curvature perturbations for $\gamma>2/3$. The
third possible future asymptotic attractor is a solution with negative
spatial curvature where the scalar field energy density remains
proportional to the curvature with $\Omega_\phi=2/\kappa^2$
($\gamma>2/3,\kappa^2>2$). We find that solutions with $\Omega_\phi=0$
are never late-time attractors.
\end{abstract}

\pacs{PACS numbers(s): 04.20.Jb, 98.80.Hw \hfill gr-qc/9901014}


\section{Introduction}

There have been a number of studies of spatially homogeneous scalar
field cosmological models with an exponential potential, with
particular emphasis on the possible existence of inflation in such
models if the potential is sufficiently flat
\cite{LucchinMatarrese,Halliwell87,BurdBarrow88,KitadaMaeda93,Coley97a}.
Although modern supergravity theories generally predict exponential
potentials that are too steep to drive inflation, these models still
have important cosmological consequences.  For example, in models with
barotropic matter, such as dust or radiation, there exist spatially
homogeneous and isotropic `scaling solutions' in which the scalar
field energy density tracks that of the perfect fluid
\cite{Wetterich88,Wands93,Copeland97,FerreiraJoyce97b}, so that a
significant fraction of the energy density of the Universe at late
times may be contained in the homogeneous scalar field whose dynamical
effects mimic the barotropic matter.  The tightest constraint on these
cosmological models comes from primordial nucleosynthesis bounds on
any such relic density during the radiation dominated era
\cite{Wetterich88}.  More recently attention has been directed to the
possible effect of such a scalar field on the growth of large-scale
structure in the universe
\cite{FerreiraJoyce97b,FerreiraJoyce97a,Wetterich95,Viana97}.

A phase-plane analysis of the spatially homogeneous and isotropic {\it zero
curvature} models \cite{Copeland97} has shown that these scaling
solutions are the unique late-time attractors whenever they exist.
The stability of these scaling solutions in more general spatially
homogeneous cosmological models was studied in \cite{Billyard98}.  In
this article we shall study scaling solutions in the general class of
spatially homogeneous and isotropic models with non-zero curvature.

The governing equations for a self-interacting scalar field with an
exponential potential energy density
\begin{equation}
V = V_0 e^{-\kappa \phi}, \label{1}
\end{equation}
where $V_0$ and $\kappa$ are positive constants, evolving in a
Robertson-Walker spacetime containing a separately conserved perfect
fluid, are given by
\begin{eqnarray}
\d{H} &=& -\frac{1}{2} (\gamma \rho_\gamma + \d{\phi}^2)-K, \label{2}\\
\d{\rho}_\gamma &=& -3 \gamma H \rho_\gamma, \label{3}\\
\dd{\phi}&=& -3H\d{\phi} + \kappa V, \label{4}
\end{eqnarray}
subject to the Friedmann constraint
\begin{equation}
H^2 = \frac{1}{3} (\rho_\gamma + \frac{1}{2} \dot{\phi}^2 + V) + K,
\label{5}
\end{equation}
where $K = -k R^{-2}$ and $k$ is a constant that can be scaled to $0$,
$\pm 1$, $H\equiv\dot{R}/R$ is the Hubble parameter, an overdot
denotes ordinary differentiation with respect to time $t$, and units
have been chosen so that $8 \pi G =1$.  In the above we have assumed
that the perfect fluid satisfies the barotropic equation of state
\begin{equation}
p_\gamma = (\gamma -1) \rho_\gamma, \label{6}
\end{equation}
where $\gamma$ is a constant which satisfies $0 < \gamma < 2$.
We also note that the total energy density of the scalar field is given by
\begin{equation}
\rho_\phi = \frac{1}{2}\dot{\phi}^2 +V. \label{7}
\end{equation}

Defining
\begin{equation}
x \equiv \frac{\dot{\phi}}{\sqrt{6}H} \q , \q y \equiv
\frac{\sqrt{V}}{\sqrt{3}H}\q , \q \Omega \equiv \frac{\rho_\gamma}{3
H^2},  \label{8}
\end{equation}
and the new logarithmic time variable $\tau$ by
\begin{equation}
\frac{d \tau}{dt} \equiv H, \label{9}
\end{equation}
equations (\ref{2}--\ref{4}) can be written as the three-dimensional
autonomous system:
\begin{eqnarray}
x' &=& -3x + \sqrt{\frac{3}{2}}\kappa y^2 + \frac{3}{2}x \left[ \left(
\gamma - \frac{2}{3} \right) \Omega + \frac{2}{3} (1 + 2x^2 -y^2)
\right], \label{10}\\
y' &=& \frac{3}{2} y \left[ -\sqrt{\frac{2}{3}} \kappa x + \left(
\gamma -\frac{2}{3}  \right) \Omega + \frac{2}{3} (1+2x^2 -y^2)
\right], \label{11}\\
\Omega' &=& 3 \Omega \left[\left(\gamma - \frac{2}{3}\right)(\Omega
-1) + \frac{2}{3} (2x^2 -y^2)  \right], \label{12}
\end{eqnarray}
where a prime denotes differentiation with respect to $\tau$, and
equation (\ref{5}) becomes
\begin{equation}
1 - \Omega - x^2 -y^2 = KH^{-2}, \label{13}
\end{equation}
where
\begin{equation}
 \q \Omega_\phi \equiv \frac{\rho_\phi}{3H^2} = x^2 + y^2. \label{14}
 \end{equation}.


\section{Qualitative Analysis}

\subsection{Invariant Sets and Monotone Functions}

The physical region of the state-space is constrained by the
requirement that the energy density be non-negative; i.e., $\Omega
\geq 0$. Furthermore, from equation (\ref{13}) we find that in the
variables used the state-space is bounded for $k=0$ and $k=-1$ but not
for $k=+1$.

Geometrically the zero-curvature models ($k=0$) are represented by the
paraboloid
\begin{equation}
\Omega +x^2 +y^2 = 1, \label{paraboloid}
\end{equation}
in the $(\Omega,x,y)$ state-space.  It is therefore possible to divide
the complete state-space up into a number of invariant sets. (See
Table \ref{invariant}.)

The existence of a monotone function in any invariant set rules out
the existence of periodic orbits, recurrent orbits, and equilibrium
points in that set and serves to determine the asymptotic behaviour in
that set \cite{WainwrightEllis,Hale}.  In the three-dimensional sets
$\cal{A}$ and $\cal{C}$ we define the function
\begin{equation}
Z=\frac{\Omega^2}{(\Omega +x^2 +y^2 - 1)^2}.\label{Z}
\end{equation}
{}From Eqns. (\ref{10}-\ref{12}) we find that
\begin{equation}
Z'=2(2-3\gamma)Z.
\end{equation}
For $\gamma >2/3$ we have that $Z$ is a monotone decreasing function
along orbits in sets $\cal{A}$ and $\cal{C}$, which therefore implies
that $\Omega\to 0$ along these orbits.  For $\gamma <2/3$ we have that
$Z$ is a monotone increasing function along orbits in sets $\cal{A}$
and $\cal{C}$ which also implies that $K/6H^2 \to 0$ along these
orbits. (The case $\gamma=2/3$ will be dealt with separately in
Sec. \ref{2c}.) Hence we observe that in the non-zero-curvature models
either the curvature approaches zero or the energy density of the fluid
approaches zero as time evolves forward.  That is, the asymptotic
behaviour of these models can be completely determined by analyzing
the zero-curvature models (see \cite{Copeland97}), and the scalar
field models with no fluid matter (see
\cite{Halliwell87,BurdBarrow88,KitadaMaeda93,Coley97a}).

\subsection{Local Stability  of the Equilibrium Points}

Due to the existence of the monotonic function Eq. (\ref{Z}), all we
need to do is locally analyze the equilibrium points of the system
(\ref{10}--\ref{12}). There are seven equilibrium points, labelled
P1--P7. The equilibrium points and their local stability are listed in
Table {\ref{singular}}.

The exact solution corresponding to the equilibrium point {\bf P1} is
the standard {\bf Milne model}.  Equilibrium points {\bf P2} and {\bf
P3} correspond to a massless scalar field models which are essentially
equivalent to {\bf stiff perfect fluid Friedmann-Robertson-Walker
models}.  The remaining equilibrium points and their corresponding
exact solutions are discussed later in the text.

\subsection{The Bifurcation Values}\label{2c}

Bifurcations of the the system (\ref{10}--\ref{12}) occur for various
values of the equation of state parameter for the perfect fluid,
$\gamma$, and the scalar field potential parameter, $\kappa$. The
bifurcation values are $\gamma=2/3$, $\kappa^2=2$, $\kappa^2=6$, and
$\kappa^2=3\gamma$.  Each of these bifurcation values will be
discussed in turn.

\subsubsection{$\gamma=2/3$}

If $\gamma=2/3$ then the equilibrium points of the system
(\ref{10}--\ref{12}) are points {\bf P2}, {\bf P3}, {\bf P4} as well
as the non-isolated lines of equilibria given by
$${\bf L1}=(\Omega = \Omega_0,x=0,y=0)$$ and 
$${\bf L2} =
(\Omega =
\Omega_s,x=\frac{\sqrt{2}}{\sqrt{3}\kappa},y=\frac{2}{\sqrt{3}\kappa})$$
where $0\leq \Omega_0,\Omega_s<\infty$. These two lines of equilibria
are the degeneracies of equilibrium points {\bf P1}, {\bf P5}, {\bf
P6} and {\bf P7}.  Fortunately the analysis of the dynamics of the
system are simplified through the observation that the function
defined by Eq. (\ref{Z}) is constant when $\gamma=2/3$.  This implies
that the dynamics of the system (\ref{10}--\ref{12}) are restricted to
level surfaces of the function $Z=Z(x,y,\Omega)$. Essentially this
implies that dynamics of the three-dimensional system
(\ref{10}--\ref{12}) can be thought of as a one-parameter family of
two-dimensional surfaces (paraboloids in this case), where the
dynamics on each surface are identical and equivalent to the dynamics
on any other surface, (including the surfaces represented by $Z=0$ and
$1/Z=0$). The future asymptotic attractor for all models is the
power-law inflationary model represented by equilibrium point {\bf P4}
if $\kappa^2<2$.  If $\kappa^2>2$ then the matter scaling solutions
represented by the line of equilibria {\bf L2} are the future
asymptotic attractors for all models.

\subsubsection{$\kappa^2=2$}

If $\kappa^2=2$ then points {\bf P4} and {\bf P5} coalesce. As
$\kappa^2\to 2^-$, {\bf P5} $\to${\bf P4}, and as $\kappa^2$ increases
past 2, the stability of one of the eigendirections changes. In
essence {\bf P4} and {\bf P5} undergo a transcritical bifurcation
\cite{Wiggins}.  The point {\bf P4} is a saddle node when
$\kappa^2=2$.  If $\kappa^2=2$ and $\gamma>2/3$ then both the negative
and zero curvature models tend to the equilibrium point {\bf P4}.  For
the positively curved models, this same point acts like a saddle with
a one-dimensional unstable manifold when $\kappa^2=2$ and
$\gamma>2/3$.  If $\kappa^2=2$ and $\gamma<2/3$ then all models are
attracted to the matter scaling solution represented by {\bf P7}.

\subsubsection{$\kappa^2=6$}

If $\kappa^2=6$, then equilibrium points {\bf P4} and {\bf P2}
coalesce.  However, the future asymptotic behaviour of the system can
be determined directly from Table \ref{singular}.

\subsubsection{$\kappa^2=3\gamma$}

If $\kappa^2=3\gamma$ then {\bf P7} and {\bf P4} coalesce.  If
$\gamma<2/3$ then this point attracts those orbits in the physical
state space, i.e., it is the future attractor.  If $\gamma>2/3$ then
there exists a one-dimensional unstable manifold. If
$\kappa^2=3\gamma$ and $\gamma>2/3$ then the negatively curved models
are attracted to the point {\bf P5}, the zero curvature models are
attracted to the point {\bf P4}.

\subsection{Zero-Curvature Models}

The zero-curvature models are contained in the two-dimensional
invariant set $\cal B \cup \cal E$. A qualitative analysis of this
plane-autonomous system is given in \cite{Copeland97}.  The well-known
{\bf power-law inflationary solution} for $\kappa^2<2$
\cite{Halliwell87,BurdBarrow88,KitadaMaeda93,Coley97a} corresponds to
the equilibrium point $\bf P4$, which is shown to be stable (i.e.,
attracting in the two-dimensional invariant set $\cal B \cup \cal E$)
for $\kappa^2 < 3 \gamma$ in the presence of a barotropic fluid.

In addition, for $0<\gamma <2$ there exists a {\bf matter scaling
solution} corresponding to the equilibrium point $\bf P7$, whenever
$\kappa^2 > 3\gamma \cite{Wetterich88}$. The equilibrium point is
stable in the two-dimensional invariant set $\cal B \cup \cal E$, (a
spiral for $\kappa^2 > 24 \gamma^2/(9\gamma-2)$, otherwise a node) so
that the corresponding cosmological solution is a late-time attractor
in the class of zero-curvature models in which neither the
scalar-field nor the perfect fluid dominates the evolution and we have
\begin{equation}
\Omega_\phi = {3\gamma \over \kappa^2} \ .
\end{equation}
The effective equation of state for the scalar field is given by
$$\gamma_\phi \equiv \frac{(\rho_\phi + p_\phi)}{ \rho_\phi} =
\frac{2x^2}{x^2 + y^2} $$
which is the same as the equation of state parameter for the perfect
fluid at this equilibrium point; i.e., $\gamma_\phi=\gamma$.  The
solution is referred to as a matter scaling solution since the energy
density of the scalar field remains proportional to that of the
barotropic perfect fluid according to $\rho_\gamma/\rho_\phi =
(\kappa^2-3\gamma)/3\gamma$ \cite{Wetterich88}.  Since this matter
scaling solution corresponds to an equilibrium point of the system
(\ref{10}--\ref{12}), we note that it is a self-similar cosmological
model \cite{WainwrightEllis}.

\subsection{Non-Zero Curvature Models}

\subsubsection{$\gamma>2/3$} 
If $\kappa^2<2$, then the future asymptotic state for the
negative-curvature models and a subset of the positive-curvature
models is the standard zero-curvature power-law inflationary solution,
represented by equilibrium point {\bf P4}, where $\Omega_\phi=1$.
Previous analysis has shown that this power-law inflationary solution
is a global attractor in spatially homogeneous models in the absence
of a perfect fluid (except for a subclass of Bianchi type IX models
which recollapse) \cite{KitadaMaeda93}. However, more recent analysis
has shown that this power-law inflationary solution is also a global
attractor in spatially homogeneous Bianchi class B models (as
classified by Ellis and MacCallum \cite{EllisMacCallum}) in the
presence of a perfect fluid \cite{billyardcoleyvandenhoogen98} with
the same restrictions on $\gamma$ and $\kappa$ as above.

If $\kappa^2>2$, then the negative-curvature models are asymptotic
towards the {\bf curvature scaling solution}, represented by
equilibrium point {\bf P5}, where
\begin{equation}
\Omega_\phi = {2\over \kappa^2} \ .
\end{equation}
Part of the motivation here is to study the stability of these scaling
solutions.  It is known that the equilibrium point {\bf P5} is the
future asymptote for the Bianchi type V and Bianchi type VII$_h$
models (see \cite{Billyard98,billyardcoleyvandenhoogen98}).

\subsubsection{$\gamma<2/3$}   

If $\kappa^2<3\gamma$, then power-law inflation, represented by
equilibrium point {\bf P4}, is again the future asymptotic state for
the negative-curvature models and a subset of the positive-curvature
models. But for $\kappa^2>3\gamma$ the matter scaling solution,
represented by equilibrium point {\bf P7}, takes over as the future
asymptotic attractor.

The linearization of system (\ref{10}--\ref{12}) about the equilibrium
point {\bf P7} yields two negative real eigenvalues and the eigenvalue
$(3 \gamma -2)$. Hence the matter scaling solution is only stable for
$\gamma < \frac{2}{3}$.  For $\gamma > \frac{2}{3}$ the equilibrium
point {\bf P7} is a saddle with a two-dimensional stable manifold
(lying in the set $\cal B \cup \cal E$ representing the zero-curvature
models) and a one-dimensional unstable manifold.


\section{Cosmological Implications}

In order to solve the flatness problem of the standard model, a period
of accelerated expansion where the the spatial curvature is driven to
zero ($K/H^2\to0$) is desirable \cite{KolbTurner}.  Power-law models
of inflation driven by scalar fields with exponential potentials with
$\kappa^2<2$ provide an interesting model of inflation where exact
analytic solutions are possible, not only for the background
homogeneous fields \cite{LucchinMatarrese}, but also for inhomogeneous
linear perturbations \cite{LythStewart}. We have shown that the
power-law inflation solution, represented by equilibrium point {\bf
P4}, with $\Omega_\phi=1$ and zero curvature, is a stable late time
attractor solution in models with spatial curvature and non-zero
density of barotropic fluid for $\gamma>2/3$ and $\kappa^2<2$ and for
$\gamma<2/3$ and $\kappa^2<3\gamma$.

Power-law inflation can also be driven by a barotropic fluid with
$\gamma<2/3$. However, we have shown that this fluid dominated
solution, corresponding to equilibrium point {\bf P6}, is never stable
for $\gamma>0$ in the presence of a scalar field with an exponential
potential. Instead, for $\gamma<2/3$ and $\kappa^2>3\gamma$ we have
shown that the matter scaling solution, corresponding to equilibrium
point {\bf P7} with zero spatial curvature, is the late time attractor
for all models with zero or negative curvature, and for a subset of
models with positive curvature.
 
This leaves a possible relic density problem
\cite{Copeland97}. Although the spatial curvature may become
negligible, the energy density in a scalar field with an exponential
potential cannot be diluted away by inflation driven by barotropic
matter unless $\gamma\to0$. In typical slow-roll inflation models
$\gamma\sim0.1$ \cite{LiddleLyth93}, and thus the exponential
potential can have a non-negligible density during and after
inflation.  During the subsequent radiation dominated era
($\gamma=4/3$) the scalar field rapidly approaches the matter scaling
solution where $\Omega_\phi=4/\kappa^2$. Models of primordial
nucleosynthesis require $\Omega_\phi<0.13$ to $0.2$
\cite{FerreiraJoyce97b}. Taking the more conservative upper limit we
can conclude that the existence of scalar fields with exponential
potentials is incompatible with standard models of nucleosynthesis for
$\kappa^2<20$, unless the scalar field has for some reason not reached
its scaling solution. Conventional inflation models with
$0<\gamma<2/3$ cannot prevent $\Omega_\phi$ reaching its equilibrium
value soon after inflation.

Observational evidence suggests we live in a Universe with a present
day energy-density $\Omega_0$ bounded between 0.1 and 0.3
\cite{ColesEllis}. An intriguing possibility is that the dominant dark
matter in the Universe could be in the form of a scaling scalar field
with an exponential potential, which would be compatible with a low
density, but spatially flat universe for $\kappa^2\sim4$. However,
this would only be compatible with the nucleosynthesis bounds quoted
above if the scalar field is far from its scaling solution at the time
of nucleosynthesis \cite{FerreiraJoyce97b}. Such a model is also
likely to be in conflict with conventional models of structure
formation as an inhomogeneous scalar field behaves differently from
the barotropic matter and scalar field gradients can exert large
pressures that resist gravitational collapse on small scales
\cite{Viana97,FerreiraJoyce97a}.

We have shown that the matter scaling solution is always unstable to
curvature perturbations in the presence of ordinary matter ($\gamma
\geq 1$); i.e., the matter scaling solution is no longer a late-time
attractor in this case. However, it does still correspond to an
equilibrium point in the governing autonomous system of ordinary
differential equations and hence there are cosmological models that
can spend an arbitrarily long time `close' to this solution.  Indeed
the curvature of the Universe is presently constrained to be small by
cosmological observations, so it is possible that the matter scaling
solution could be important in the description of our actual
Universe. Negative curvature spaces are interesting since they can
support a variety of possible topologies, leading to, for example,
hyperbolic null geodesic motions which can produce interesting
patterns in the cosmic microwave background radiation
\cite{BarrowLevin}.  At late times, all models with $\kappa^2>2$,
$\gamma>2/3$ and negative spatial curvature approach the curvature
scaling solution, corresponding to equilibrium point {\bf P5}.

Classical cosmological tests such as number counts, or the luminosity of
standard candles, should in principle be able to distinguish between
the different possible low-density models.
The deceleration parameter at any point in the phase-space is given by
\begin{equation}
q_0 \equiv - {\ddot{R}R \over \dot{R}^2}
 = \left( {3\gamma\over2} - 1 \right)\Omega + 2x^2 - y^2 \ .
\end{equation}
This reduces to the standard result $q_0=3\gamma/2-1$ for the matter
scaling solution, or $q_0=0$ for the curvature scaling solution. The
dynamics of the cosmological scale factor in these these solutions is
indistinguishable from the matter dominated or curvature dominated
solutions. However the luminosity distance as a function of redshift
depends not only upon the evolution of the scale factor, but also upon
the spatial geometry, and hence the curvature scaling solution can in
principle be distinguished from the curvature dominated Milne model.

Recent results from high redshift supernovae searches suggest that the
deceleration parameter is negative at present; i.e., the Universe is
in fact accelerating \cite{Garnavich}. Although it is possible to
obtain $q_0<0$ when $y^2>2x^2$, we have found from numerical
experimentation that this is unlikely during a transition from matter
scaling to curvature scaling regime when $\kappa^2>2$. Instead it
would only seem to be compatible with either barotropic matter with
$\gamma<2/3$ or a scalar field dominated solution, corresponding to
equilibrium point {\bf P4}, with $\kappa^2<2$.


\section{Conclusions}

A complete qualitative analysis of the dynamical system describing the
evolution of spatially homogeneous and isotropic models containing a
non-interacting perfect fluid and a scalar field with an exponential
potential has yielded the following results:
\begin{enumerate}
\item  The past asymptotic
state for all expanding models is the massless scalar
field solution.  
\item The future asymptotic state depends upon the values of
the equation of state parameter for the fluid, $\gamma$, and the
steepness of the potential $\kappa$.  The future asymptotic state is either
\begin{enumerate}
\item the standard power-law inflationary solution (if $\gamma<2/3$,
$\kappa^2<3\gamma$ or $\gamma>2/3$, $\kappa^2<2$),
\item a matter
scaling solution (if $\gamma <2/3$, $\kappa^2>3\gamma$), or 
\item in the case of the negative-curvature models, a curvature scaling
model with no barotropic matter ($\gamma>2/3$, $\kappa^2>2$).
\end{enumerate}
\item In non-zero curvature models with ordinary matter, (i.e., $\gamma
\geq1$), the matter scaling solution is never stable.
\end{enumerate}

Current cosmological observations seem to indicate $\Omega<1$ in
ordinary matter which could be due to the presence of (negative)
spatial curvature, and or some other form of energy density such as a
scalar field with exponential potential. It is known that the matter
scaling solutions are late-time attractors (i.e., stable) in the
subclass of zero-curvature isotropic models \cite{Copeland97}.  It is
also known that the matter scaling solutions are stable (to shear and
curvature perturbations) in generic anisotropic Bianchi models when
$\gamma<2/3$ \cite{Billyard98}.  However, when $\gamma>2/3$, and
particularly for ordinary matter with $\gamma \ge 1$, the matter
scaling solutions are unstable; essentially they are unstable to
curvature perturbations, although they are stable to shear
perturbations \cite{Billyard98}.  Even though it is unstable, this
matter scaling solution may still play an important role in describing
the intermediate (or transient) physics between the past and future
asymptotic attractors.

The negative curvature scaling solutions may also be of importance
since they are self-similar cosmological models corresponding to an
equilibrium point of the dynamical system (\ref{10}--\ref{12}) where
the scalar field still has a non-vanishing contribution to the energy
density.  Cosmological tests such as the magnitude-redshift relation
for standard candles or detailed observations of the microwave
background should in principle be able to determine whether these
solutions could describe the present state of the universe.

\acknowledgements 
RvdH and AAC were supported by the Natural Sciences
and Engineering Research Council of Canada. RvdH was also supported by
a grant through the St.F.X. University Council on Research.  RvdH
would like to thank Colleen Meagher and Bill MacMillan for checking
the analysis and doing some numerical calculations.


\begin{table}\squeezetable
\caption{Invariant sets for the dynamical system given by equations
(\ref{10}--\ref{12}).}
\label{invariant}
\begin{tabular}{lccl}
Label       & Set $(\Omega,x,y)$                     & Dimension &
Description \\
\hline
$\cal{A}$\q & $\Omega>0$ and $\Omega +x^2 +y^2 > 1$\q  &    3 \q &
positive curvature, non-vacuum\\
$\cal{B}$   & $\Omega>0$ and $\Omega +x^2 +y^2 = 1$  &    2  &  zero
curvature, non-vacuum\\
$\cal{C}$   & $\Omega>0$ and $\Omega +x^2 +y^2 < 1$  &    3  &
negative curvature, non-vacuum\\
$\cal{D}$   & $\Omega=0$ and $\Omega +x^2 +y^2 > 1$  &    2  &
positive curvature, no fluid matter\\
$\cal{E}$   & $\Omega=0$ and $\Omega +x^2 +y^2 = 1$  &    1  &  zero
curvature, no fluid matter\\
$\cal{F}$   & $\Omega=0$ and $\Omega +x^2 +y^2 < 1$  &    2  &
negative curvature, no fluid matter
\end{tabular}

\end{table} 

\begin{table}\squeezetable
        \caption{Equilibrium points of the dynamical system
        (\ref{10}--\ref{12}) and values of the parameters $\gamma$ and
        $\kappa$ for which the equilibrium point is a future
        attractor.}  \label{singular} 
        \begin{tabular}{|c|c|c|c|c|c|c|} Label& $\Omega$ & $x$ & $y$ &
        $^3R$ & Eigenvalues & Stability Conditions \\ \hline & & & & &
        $2-3\gamma$ & \\ P1 & 0 & 0 & 0 & -1 & -2 & Never\\ & & & & &
        1 & \\ \hline & & & & & $6-3\gamma$ & \\ P2 & 0 & 1 & 0 & 0 &
        4 & Never\\ & & & & & $3 - \frac{\sqrt{6}}{2}\kappa$ & \\
        \hline & & & & & $6-3\gamma$ & \\ P3 & 0 & -1 & 0 & 0 & 4 &
        Never\\ & & & & & $3 + \frac{\sqrt{6}}{2}\kappa$ & \\ \hline &
        & & & & $\kappa^2 - 3\gamma$ & $\gamma< \frac{2}{3},\ \
        \kappa^2 < 3\gamma$ \\ P4 & 0 &
        $\displaystyle\frac{\kappa}{\sqrt{6}}$ &
        $\displaystyle\sqrt{1-\frac{\kappa^2}{6}}$ & 0 & $\kappa^2-2$
        & and \\ & & & & & $\frac{1}{2}\left(\kappa^2 - 6 \right)$ &
        $\gamma > \frac{2}{3},\ \ \kappa^2 < 2$ \\ \hline & & & & &
        $2-3\gamma$ & \\ P5 & 0 &
        $\displaystyle\frac{\sqrt{2}}{\sqrt{3}\kappa}$ &
        $\displaystyle\frac{2}{\sqrt{3}\kappa}$ & $ \displaystyle
        \frac{2-\kappa^2}{2}$ & $-1 \pm \sqrt {\frac{8}{\kappa^2}-3}$
        & $\gamma>\frac{2}{3}, \ \ \kappa^2 > 2$ \\ & \ & & & & & \\
        \hline & & & & & $3\gamma -2$ & \\ P6 & 1 & 0 & 0 & 0 &
        $\frac{1}{2}\left(3\gamma-6 \right)$ & Never \\ & & & & &
        $\frac{3}{2}\gamma$ & \\ \hline & & & & & $3\gamma-2$ & \\ P7
        & $\displaystyle \frac{\kappa^2-3\gamma}{\kappa^2}$ &
        $\displaystyle \frac{\gamma\sqrt{6}}{2\kappa}$ &
        $\displaystyle \frac{\sqrt{12\gamma-6\gamma^2}}{2\kappa}$ & 0
        & $\frac{3}{4}\left[ (\gamma-2) \pm \sqrt{(\gamma-2)^2 +
        \frac{8\gamma}{\kappa^2}(\gamma-2) \left(\kappa^2-3\gamma
        \right)}\right]$ & $\gamma<\frac{2}{3}, \ \
        \kappa^2>3\gamma$\\ & \ & & & & & \\ \end{tabular}
\end{table}
\end{document}